\documentclass[pra,aps,twocolumn,showpacs]{revtex4}
\usepackage{amsmath,amsfonts,amssymb,graphics,graphicx,epsfig,color,times}
\usepackage[latin1]{inputenc}
 
\graphicspath{{Bilder/}}
 
\newcommand{\ket}[1]{\left| #1 \right.\rangle}

\newcommand{\ad}{\hat{a}^\dagger}
\renewcommand{\a}{\hat{a}}
\newcommand{\nb}{\hat{n}}

\DeclareMathSymbol{\theta}{\mathord}{letters}{"23}
\DeclareMathSymbol{\rho}{\mathord}{letters}{"25}
\DeclareMathSymbol{\phi}{\mathord}{letters}{"27}
 
\DeclareMathSymbol{\vartheta}{\mathord}{letters}{"12}
\DeclareMathSymbol{\varphi}{\mathord}{letters}{"1E}
\DeclareMathSymbol{\varrho}{\mathord}{letters}{"1A}
\begin{document}
 
\newcommand{\nn}{{\mathbbm{N}}}
\newcommand{\rr}{{\mathbbm{R}}}
\newcommand{\cc}{{\mathbbm{C}}}
\newcommand{\id}{{\sf 1 \hspace{-0.3ex} \rule{0.1ex}{1.52ex}\rule[-.01ex]{0.3ex}{0.1ex}}}
\newcommand{\me}{\mathrm{e}}
\newcommand{\mi}{\mathrm{i}}
\newcommand{\md}{\mathrm{d}}
\renewcommand{\vec}[1]{\text{\boldmath$#1$}}

\title{Ultracold bosons in disordered superlattices: Mott-insulators induced 
 by tunneling}
\author{D. Muth, A.\ Mering and M.\ Fleischhauer}
\affiliation{Fachbereich Physik, Technische Universit\"at Kaiserslautern, D-67663 Kaiserslautern, Germany}

\begin{abstract}
We analyse the phase diagram of ultra-cold bosons in a one-dimensional superlattice potential with disorder using the \textit{time 
evolving block decimation}  algorithm for infinite sized systems (iTEBD). For degenerate potential energies within the unit cell
of the superlattice loophole-shaped insulating phases with non-integer filling emerge with a particle-hole gap proportional to
the boson hopping. Adding a small amount of disorder destroys this gap. For not too large disorder the loophole Mott regions 
detach from the axis of vanishing hopping giving rise to insulating islands. Thus the system shows a 
transition from a compressible Bose-glass to a Mott-insulating phase with \textit{increasing} 
hopping amplitude. We present a straight forward effective model for the dynamics within a unit cell which 
provides a simple explanation for the emergence of Mott-insulating islands. In particular it gives rather accurate predictions for the inner 
critical point of the Bose-glass to Mott-insulator transition.
\end{abstract}
 
 
\date{\today}
 
\maketitle

\section{Introduction}

Ultra-cold atomic gases in light induced periodic potentials have become an important experimental testing ground
for concepts of solid-state and many-body physics since they allow the realization of precisely controllable 
model Hamiltonians with widely tunable parameters. This development was triggered by the theoretical proposal
of Jaksch {\it et al.} \cite{lit:Jaksch-PRL-1998} that ultra-cold bosonic atoms in an optical lattice are well described by the 
Bose-Hubbard model and the subsequent observation of the superfluid-Mott insulator transition in that system
by Greiner {\it et al.} \cite{lit:Greiner-Nature-2002}. A characteristic feature of light induced periodic potentials is the 
possibility to modify their properties in a simple way. E.g. when phase locked lasers with different but commensurable
frequencies are superimposed to create a periodic dipole potential, different types of superlattices with more
complex unit cells can be constructed \cite{lit:Jaksch-PRL-1998, lit:santos_prl_2004,  lit:Peil-PRA-2003,lit:Roth-PRA-2003,lit:Buonsante-PRA-2004a,lit:Buonsante-PRA-2004b,lit:Buonsante-PRA-2005,lit:Rousseau-PRB-2006}.
Superimposed optical lattices with non-commensurate frequencies furthermore mimic a disordered potential
\cite{lit:inguscio_prl_2007}.

In the present paper we study the phase diagram of ultra-cold bosons in a  one-dimensional superlattice potential
with degenerate potential energies and/or degenerate tunneling rates within the unit cell. For such a system loophole shaped
Mott-insulator domains with fractional filling have been predicted by Buonsante, Penna and Vezzani \cite{lit:Buonsante-PRA-2004a} within 
a multiple-site mean-field approach, as well as with a cell strong-coupling perturbation approach \cite{lit:Buonsante-PRA-2005}. 
In contrast to the Mott lobes at integer filling known from the simple Bose-Hubbard model, which exist also in the superlattice,
the characteristic feature of the loophole insulators is a particle-hole gap that vanishes at zero boson hopping $J$. So in the
$\mu - J$ phase diagram, where $\mu$ is the chemical potential, these domains touch the $J=0$ line only in a single point.
We here perform numerical simulations using the time evolving block decimation algorithm (TEBD) introduced by Vidal
{\it et al.} \cite{vidal_part1} in the infinite system variant \cite{vidal_itebd} as well as density matrix renormalization
group (DMRG) calculations \cite{lit:Schollwoeck-RMP-2005} to determine the boundaries of the different Mott-insulating
regions in the phase diagram. We also present a simple effective model that provides a straight forward explanation for the emergence
of the loophole insulators by taking into account hopping processes between the sites of degenerate potential energy within 
a unit cell but neglecting tunneling between different unit cells. 

We then study the influence of some additional disorder potential with continuous, bounded distribution.
If the maximum amplitude of the disorder is not too large, the Mott lobes shrink in a similar way as for the simple Bose-Hubbard model
\cite{lit:Fisher-PRB-1988}.  Also the loophole Mott domains shrink. As a consequence near the critical (fractional) filling Mott-insulating islands emerge surrounded by a Bose-glass phase. A rather peculiar property of this system is the phase transition from a compressible (Bose-glass) phase for small values of the bosonic hopping $J$ to an incompressible Mott phase for larger tunneling rates, i.e. we have a tunneling 
induced Mott insulator. The effective model describing the full dynamics within a unit cell provides a simple explanation for this
and gives good quantitative predictions for the critical value $J_c$ for the compressible-to-Mott transition. The analytical predictions
are compared to numerical simulations again using the iTEBD algorithm for a superlattice with disorder.

\section{The model}\label{model}
 
We consider ultracold bosonic atoms in an optical superlattice, having a periodic structure 
with a period $l$ of some lattice sites. As shown in \cite{lit:Jaksch-PRL-1998}, the physics of these 
atoms can be described by the so called superlattice \emph{Bose-Hubbard model}, extensively 
studied in \cite{lit:Buonsante-PRA-2004b, lit:Buonsante-PRA-2005}. We here work mostly in the grand canonical BHM, 
only the calculations of the shape of the loophole insulators in sections \ref{sec:two-site_without} and 
\ref{sec:two-site} will be performed for fixed particle numbers. In second quantisation, the BHM reads as
\begin{eqnarray}\label{eq:BHM}
 \hat{H} &=&-J\sum_j t_{j}\ \left(\ad_j\a_{j+1}+\ad_{j+1}\a_{j}\right)+\frac{U}{2}\sum_j\nb_j\left(\nb_j-1\right)  \nonumber\\
&& -\sum_j (\mu-v_j)\nb_j,\label{eq:hamiltonian}
\end{eqnarray}
where $\a_j$ and $\ad_j$ are the annihilation and creation operators of the bosons at lattice site
$j$, and $\nb_j=\ad_j\a_j$ is the corresponding number operator. The particles can tunnel from one lattice site
to a neighbouring one with hopping rate $J$, $t_j$ accounts for the variation of the hopping due to the superlattice potential. $v_j$ accounts for local
variations of the potential energy within a unit cell, and $\mu$ is the (global) chemical potential.

A particularly interesting situation arises if there is a degeneracy in the tunneling amplitudes and 
local potentials within the unit cell. This will be studied in detail in the following. For simplicity 
we focus on a special superlattice structure in which only the local potential is varied with period three, namely 
\begin{equation}
 \vec v = \{v_1,v_1,v_2\}
\end{equation}
with $v_1-U<v_2<v_1<U$ and $\vec t=\{1,1,1\}$. It should be noted, that the results obtained for this case are 
qualitatively identical to the more general case $\vec t=\{t_1,t_1, t_2\}$ 
and $\vec v=\{v_1,v_1,v_2\}$. 
 
\section{superlattice without disorder}

Let us consider first a superlattice  BHM without disorder in the two cases $\vec v=\{\frac U2, \frac U2, 0\}$, 
$\vec t = \{1, 1, 1\}$, and $\vec v = \{0, 0\}$, $\vec t=\{1, 0.2\}$. As shown in Ref. \cite{lit:Buonsante-PRA-2005} within a
supercell mean-field approach such superlattices lead to loophole shaped insulator phases at fractional bosonic filling.
In the following we will determine the boundaries of these loophole phases numerically and compare them to the
mean-field predictions. Furthermore we will present a rather simple model which provides an intuitive explanation.
 
\subsection{Numerical results}\label{numerics}
\begin{figure}
 \centering
 \epsfig{file=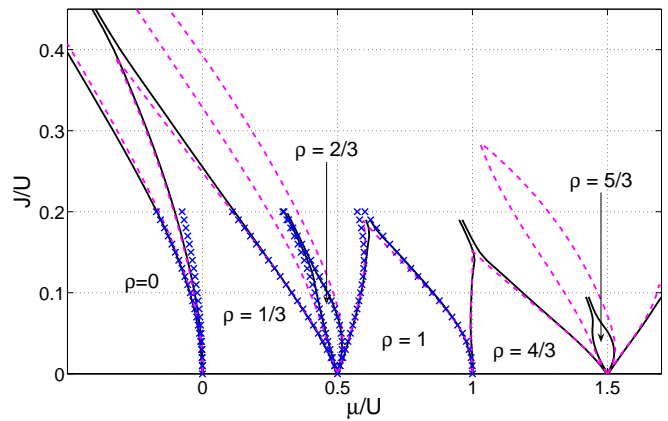,width=\columnwidth}
 \caption{(Color online) Incompressible phases for an $l=3$ superlattice with $\vec{v}=\{\frac{U}{2}, \frac{U}{2}, 0\}$ and 
$\vec{t} = \{1, 1, 1\}$ without disorder. Solid line: iTEBD; crosses: DMRG; dashed line: CSCPE from \cite{lit:Buonsante-PRA-2005}. 
Simulation parameter for the iTEBD are $\chi = 5$, $D-1 = 3$, $\beta/U = 1000$ (see appendix for definitions). 
DMRG results are obtained from a infinite size extrapolation.}
 \label{fig:phase_diagram_nodisorder}
\end{figure}

\begin{figure}
 \centering
 \epsfig{file=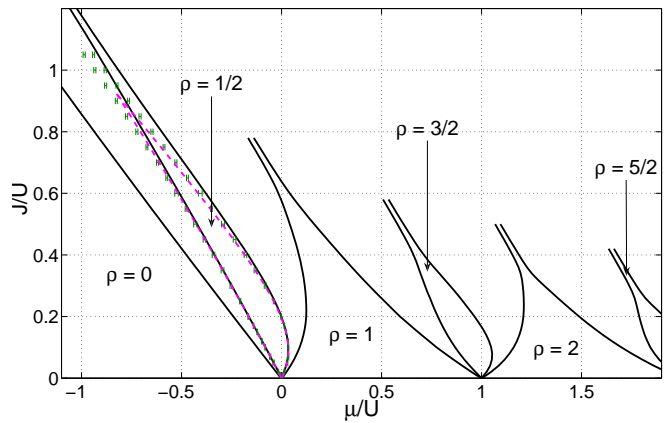,width=\columnwidth}
 \caption{(Color online) Incompressible phases for an $l=2$ superlattice with $\vec{v}=\{0, 0\}$ and $\vec{t} = \{1, 0.2\}$ 
without disorder. Solid line: iTEBD; crosses with error bars: QMC (only for $\rho = \frac12$) and dashed line: CSCPE (only for $\rho = \frac12$) both from \cite{lit:Buonsante-PRA-2005}. Simulation 
parameter for the iTEBD are $\chi = 7$, $D-1 = 4$ for $\rho < 2$ respectively $D-1 = 5$ for $\rho \ge 2$, $\beta/U = 1000$ (see appendix for definitions). For some phase boundaries $\varepsilon$ was set to $0.02$ instead of $0.005$ in order to avoid numerical artifacts.}
 \label{fig:phase_diagram_hopping}
\end{figure}

In order to calculate the boundaries of the Mott phases for the superlattice Bose Hubbard model, we 
apply the iTEBD algorithm as described in the appendix to calculate the ground state of Hamiltonian 
(\ref{eq:BHM}). We are able to calculate properties such as the local density $\rho = \overline{\langle n\rangle}$ as a function of $J$ and $\mu$. 
Using this method to map out the shape of the insulating regions we make use of the fact that for any Mott phase 
the average local density is an exact multiple of $1/l$, $l$ being the period of the superlattice. Thus the phase 
boundaries can be well approximated by the line where
\begin{equation}\label{eq:epsilon}
\mu_{\rm cr}=\mu(\overline{\langle n\rangle}),\qquad\textrm{where}\quad \overline{\langle n\rangle} = m/l \pm \varepsilon
\end{equation}
for some $m \in \mathbb{N}$ (indicating the order of the lobe) and some $1/l \gg \varepsilon > 0$ \footnote{Throughout the 
paper we set $\varepsilon = 0.005$}. This line can be calculated by finding the value of $\mu$ using a bisection method 
for a set of given $m$ and $J$. Figures \ref{fig:phase_diagram_nodisorder} and \ref{fig:phase_diagram_hopping} show the 
results of this approach for the two different superlattice potentials 
specified above.


The phase diagrams consist of a number of incompressible Mott phases, separated by a superfluid region. In contrast to the
BHM for a simple lattice, there are however two types of insulating phases: The lobe-shaped ones, well known from the simple-lattice 
Bose-Hubbard model which have a finite extent at $J = 0$ and the loophole-shaped ones which vanish at $J=0$. In general there are 
$l$ distinct insulating regions for a superlattice of period $l$. A loophole is present, whenever the local 
potential $v_j$ is the same for two sites in the same unit cell. The following section will give a qualitative understanding of 
these loophole Mott regions.

Figure \ref{fig:phase_diagram_nodisorder} shows our results in the case $\vec v=\{\frac U2, \frac U2, 0\}$, $\vec t = \{1, 1, 1\}$, together with the cell strong coupling pertubative expansion (CSCPE) results from \cite{lit:Buonsante-PRA-2005}. As a check of our numerics we added numerical results from a density matrix renormalisation group (DMRG) calculation \cite{lit:Schollwoeck-RMP-2005}. The existence of non integer insulating phases at $J=0$ in this special superlattice has a direct connection to the case of a binary disorder-BHM \cite{lit:Mering-2}, which arises for example in the presence of a second,
immobile particle species (here of filling $\frac 13$ with an inter species interaction of $v=- \frac U2$).\\
 
Figure \ref{fig:phase_diagram_hopping} shows the numerical results together with the quantum Monte Carlo (QMC) results and the CSCPE data from \cite{lit:Buonsante-PRA-2005}. The agreement between our numerics and the CSCPE is naturally good for small $J$ but deteriorates for larger $J$.
It is also apparent that while the insulator lobes are rather well described by the CSCPE approach, it is much less accurate for the
loophole insulator regions, in particular for the case of varying potential depth (see figure \ref{fig:phase_diagram_nodisorder}).
 
\subsection{Two-site model}\label{sec:two-site_without}

We will argue in the following that the loophole insulator phases can entirely be understood from the
effective dynamics within a unit cell of the superlattice. To this end let us discuss the above situation, where $\vec v=\{v_1, v_1, v_2\}$,  with $v_2<v_1$. 

The presence of Mott lobes at fractional filling with a finite extend at $J=0$ can easily be understood along the lines of the
simple-lattice BHM. As long as the filling is less than $\frac 13$ the particles will occupy sites with local potential $v_2$. Thus the chemical
potential reads
\begin{equation}
 \mu_{\frac13}^-=v_2.
\end{equation}
When the filling reaches the value $\frac 13$ additional particles will start to occupy sites with local potential $v_1$, giving rise to a particle-hole gap
\begin{equation}
\Delta\mu_{\frac 13}= \mu_{\frac13}^+-\mu_{\frac 13}^-=v_1-v_2.
\end{equation}

To explain the existence of the loop-hole insulators, one has to take into account a finite hopping $J$. For $J=0$ any particle added to the system
between $\varrho =\frac{1}{3}$ and $\varrho = 1$ increases the total energy by the same amount $v_1$. Thus the chemical potential stays the
same. This picture changes however when a small but finite tunneling is included. If the filling exceeds the value $\frac 13$ additional particles
experience an effective superlattice potential ${\vec v}=\{v_1,v_1,U+v_2\}$ where the last term results from the interaction with particles 
already occupying sites with $v_2$. If $U> v_1-v_2$ the superlattice effectively separates into degenerate double-well problems each corresponding
to a unit cell. Due to the degeneracy of the double well any small tunneling $J$ within the unit cell of the lattice needs to be taken into
account while intra-cell tunneling can be ignored. A finite tunneling lifts the degeneracy of the single-particle states within the unit cell and leads to a splitting between symmetric and antisymmetric superpositions proportional to $J$. As long as the filling is less than $\varrho =\frac{2}{3}$ the particles occupy all sites with the smallest local potential $v_2$ and the symmetric superposition of the double-well $\{v_1,v_1\}$. After that additional particles have to go either to an already occupied side with potential $v_2$, which is however suppressed by the large repulsive particle-particle
interaction, or to the anti-symmetric superposition state. The latter requires an energy on the order of $v_1 + J$, thus leading to
another particle-hole gap on the order of $J$ induced by intra-cell tunneling. More quantitatively the gap can be calculated by diagonalizing
the two-site  Hamiltonians $\mathcal H(N_B)$ for zero, one or two particles, (i.e. $N_B=0, 1, 2$), which read
\begin{gather}
 \mathcal H(0)=0\\
 \intertext{in the basis $\{\ket{00}\}$,}
 \mathcal H(1)= \left[\begin{matrix} v_1 & -J\\-J&v_1\end{matrix}\right] \label{eq:H1}\\
  \intertext{in the basis $\{\ket{10},\ket{01}\}$, and}
 \mathcal H(2)=\left[\begin{matrix} U+2v_1 &-J\sqrt2&0\\-J\sqrt2&2v_1&-J\sqrt2\\0&-J\sqrt2&U+2v_1  \end{matrix}\right]
 \end{gather} 
in the basis $\{\ket{20},\ket{11}\ket{02}\}$. The resulting ground state energies are given by
\begin{gather}
  E(0)=0,\\
  E(1)=v_1-J,\\
  E(2)=\frac12\left(U +4v_1-\sqrt{16  J^2 + U^2}\right).
 \end{gather}
 Calculating the chemical potentials  $\mu^+_\frac23=E(2)-E(1)$ and $\mu^-_\frac23=E(1)-E(0)$
yields:
 \begin{align}
 \mu^-_\frac23&=v_1-J\\
  \mu^+_\frac23&=J+v_1+\frac U2-\frac12\sqrt{16 J^2+U^2}\\
  &= v_1 + J + \mathcal O(J^2),
 \end{align}
giving rise to a particle-hole gap
\begin{equation}\label{eq:gap}
\Delta \mu_{\frac 23} = J +{\cal O}(J^2).
\end{equation}
 A generalisation of this discussion to the case of higher order loophole insulators or larger supercells is straight forward. 
For higher order loopholes, the accuracy becomes better, since the difference in the chemical potential between 
the rightmost and the other sites scales as $U(n-1)$, where $n = \lfloor\frac{m}{l}\rfloor+1$ is the number of particles of 
the corresponding Mott-insulating lobe. This means, that the effective two-site model gets even better for higher fillings. Qualitatively, one can understand the change in the shape of the loopholes for higher order (also see figure \ref{fig:loops}) just by considering the replacement $J \rightarrow (n+1)J$ in (\ref{eq:H1}), because the single particle matrix $\mathcal H(1)$ is the only one relevant for the linear part of (\ref{eq:gap}) and because this replacement is the only influence of the other bosons already filling the lattice on the hopping in $\mathcal H(1)$.

\section{superlattice with disorder}

We now include a small disorder to the superlattice Bose Hubbard model. Of particular
interest is the effect of the disorder to the loophole insulator phases. 
Disorder can be incorporated into the model by replacing the last part of (\ref{eq:hamiltonian})
according to:
\begin{equation}
 -\sum_j (\mu-v_j)\hat n_j \rightarrow -\sum_j (\mu-v_j + \Delta_j)\hat n_j
\end{equation}
with $\Delta_j$ being independent random numbers with continuous and bounded distribution,
$\Delta_j\in [-\Delta,\Delta]$. 
In the following we will restrict our analysis to the case of the superlattice potential 
$\vec v=\{\frac U2, \frac U2, 0\}$, $\vec t = \{1, 1, 1\}$ and consider a canonical ensemble.\\
 
\subsection{Two-site model with disorder}\label{sec:two-site}

If the disorder is small, i.e. if $3\Delta < U+2v_2-v_1$, the properties of the system
in the vicinity of the $\Delta=0$ loophole insulators can again be understood by considering the
unit cell only, i.e. within an effective two-site model. The $\Delta=0$ loophole insulator can be characterized
by the number of particles per unit cell $n$ and the disorder-modified chemical potential $\mu=\{v_1+\Delta_1,v_1+\Delta_2\}$.

Defining the total local energy in one unit cell as
\begin{gather}
 T_{n_1,n_2}:= \frac U2 n_1(n_1-1)+\frac U2 n_2(n_2-1)+\Delta_1 n_1+\Delta_2 n_2,
\end{gather}
the two-site Hamiltonians can be written as
\begin{equation}
 \mathcal H_{n}(2n-2)=T_{n-1,n-1},
\end{equation}
 in the basis $\{\ket{n-1,n-1}\}$ for one particle less,
\begin{equation}
 \mathcal H_{n}(2n-1)=\left[\begin{matrix}T_{n-1,n} & -n J\\ -n J& T_{n,n-1}\end{matrix}\right],
\end{equation}
in the basis $\{\ket{n-1,n},\ket{n,n-1}\}$ for zero extra particles , and
\begin{eqnarray}
&& \mathcal H_{n}(2n)=\\
&&=\left[\begin{matrix} T_{n+1,n-1}&-J\sqrt {n(n+1)}&0\\-J\sqrt {n(n+1)}&T_{n,n}&-J
\sqrt {n(n+1)}\\0&-J\sqrt {n(n+1)}&T_{n-1,n+1}\end{matrix}\right]\nonumber
\end{eqnarray}
in the basis $\{\ket{n+1,n-1},\ket{n,n},\ket{n-1,n+1}\}$ for one additional particle.\\
 
Calculating the chemical potentials one has to keep in mind, that the breaking up of the Mott-insulator 
is determined by the smallest particle-hole excitation throughout the whole system. Since all unit cells are decoupled, 
one has to find the disorder configuration which minimizes the energy gap. Therefore one 
has to calculate 
\begin{gather}
\mu^+_{\frac{3n-1}{3}}=\underset{\Delta_1,\Delta_2}{\rm min}\left[ E_{2n}(\Delta_1,\Delta_2)-E_{2n-1}(\Delta_1,\Delta_2)\right],\label{eq:maximize}\\
 \mu^-_{\frac{3n-1}{3}}=\underset{\Delta_1,\Delta_2}{\rm max}\left[ E_{2n-1}(\Delta_1,\Delta_2)-E_{2n-2}(\Delta_1,\Delta_2)\right].\label{eq:minimize}
 \end{gather}
Since the disorder has only an effect on the local energy, we can set $\Delta_1=\Delta_2=\overline{\Delta}$.  The corresponding energies are given by
 \begin{gather}
  E_{2n-2}(\overline{\Delta},\overline{\Delta})=(n-1)(2\overline{\Delta}+U(n-2)),\\
  E_{2n-1}(\overline{\Delta},\overline{\Delta})= U(n-1)2+\overline{\Delta}(2n-1)-nJ,\\
  E_{2n}(\overline{\Delta},\overline{\Delta})=\frac12\Bigl(U(n-1)2+n(U+4\overline{\Delta})\notag\\
  \hspace{2cm}-\sqrt{8J^2(n+1)n+U^2}\Bigr).
\end{gather}
Minimization (maximization) of expressions (\ref{eq:maximize}) and (\ref{eq:minimize}) yields
\begin{gather}
\mu^-_{\frac{3n-1}{3}}=-Jn+(n-1) U + \Delta,\\
 \mu^+_{\frac{3n-1}{3}}=Jn-\frac U2+Un-\Delta\notag\\
 \hspace{2cm}-\frac12\sqrt{8J^2(n+1)n+U^2}.
\end{gather}
In order to calculate the critical tunneling rate at which the loophole insulator emerges, one needs
to solve the equation
\begin{equation}
 \mu^-_{\frac{3n-1}{3}}= \mu^+_{\frac{3n-1}{3}}.
\end{equation}
The solution can easily be found and reads
\begin{equation}\label{eq:nis1}
 J_1=\Delta\frac{U-2\Delta}{U-4\Delta}
\end{equation}
for $n=1$, and
\begin{gather}
 J_n=-\frac{1}{2n(n-1)}\Bigl[(U-4\Delta)n        \notag\\
 \hspace{1cm}-\sqrt{n^2(U^2-4U\Delta+8\Delta^2)-4\Delta(U-2\Delta)n}\Bigr].\label{eq:ngt1}
\end{gather}
for $n>1$. In both cases, the leading terms are given by
\begin{equation}\label{eq:nlin}
 J_n(\Delta)= \frac 1n\Delta+\frac{n+1}{n^2}\Delta^2+\mathcal O(\Delta^3),
\end{equation}
showing that the loophole decouples from the $J=0$-axis in the presence of disorder, resulting in an insulating island.
It should be noted that the two-site model cannot be used to calculate the maximum value of $J$ for which the loophole
insulator exists since for the vanishing of the gap at large $J$ values also inter-cell tunneling processes need to be
taken into account.

\subsection{Numerical results}\label{numerics2}
As seen above from the effective two-site model, the loophole insulator regions are decoupled from the $J=0$ axis giving incompressible islands. Although the system is for any given disorder realization not translational invariant, the iTEBD method can be used also in this case. To this end we define \emph{supercells}  each of which with the same disorder. The supercells have to be large enough such that effects from spatial correlations and finite size can be ignored. We have chosen a supercell length of 96. Increasing this length did not show any noticeable changes. 
To calculate the physical quantities, the numerical results have to be averaged over a number of different disorder realisations, namely over different sets of disorder $\vec \Delta = \{ \Delta_1, \Delta_2, \dots, \Delta_l \}$ with $\Delta_j\in[-\Delta,\Delta]$, where a boxed disorder distribution is assumed. The length of the vector $\vec \Delta$ is the same as the size of the system simulated, see appendix.
It turns out that 20 realizations provide sufficient convergence for the purpose of this paper.

 \begin{figure}
 \centering
 \epsfig{file=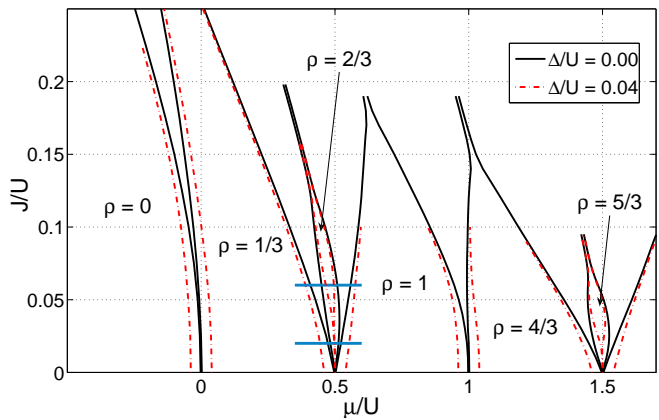,width=\columnwidth}
 \caption{(Color online) iTEBD results for boundaries of incompressible phases for $l=3$ superlattice with $\vec{v}=\{\frac{U}{2}, \frac{U}{2}, 0\}$ 
and $\vec{t} = \{1, 1, 1\}$ and a disorder amplitude $\Delta/U=0.04$. Dashed line: pure case, solid line: disordered case. For simulation 
parameters see figure \ref{fig:phase_diagram_nodisorder}. Horizontal lines indicate the positions of density cuts of figure \ref{fig:dichte}.}
 \label{fig:phase_diagram}
\end{figure}

Figure \ref{fig:phase_diagram} shows the results of the iTEBD calculations both in the pure ($\Delta=0$) and the disordered ($\Delta=0.04U$) case. The first thing to notice is the shrinking of the Mott-insulating lobes for $J=0$ due to the disorder. As known from the BHM \cite{lit:Fisher-PRB-1988,lit:Freericks-PRB-1996} the Mott-lobes shrink by an amount of $2\Delta$ at the $J=0$ axis. The second and more important thing to notice 
is the decoupling of the loophole insulator from the $J=0$ axis, meaning that there is no insulating phase for the respective filling for $J<J_{\rm crit}$, in full agreement with equations  (\ref{eq:nis1}, \ref{eq:ngt1}).\\
 
\begin{figure}
 \centering
 \epsfig{file=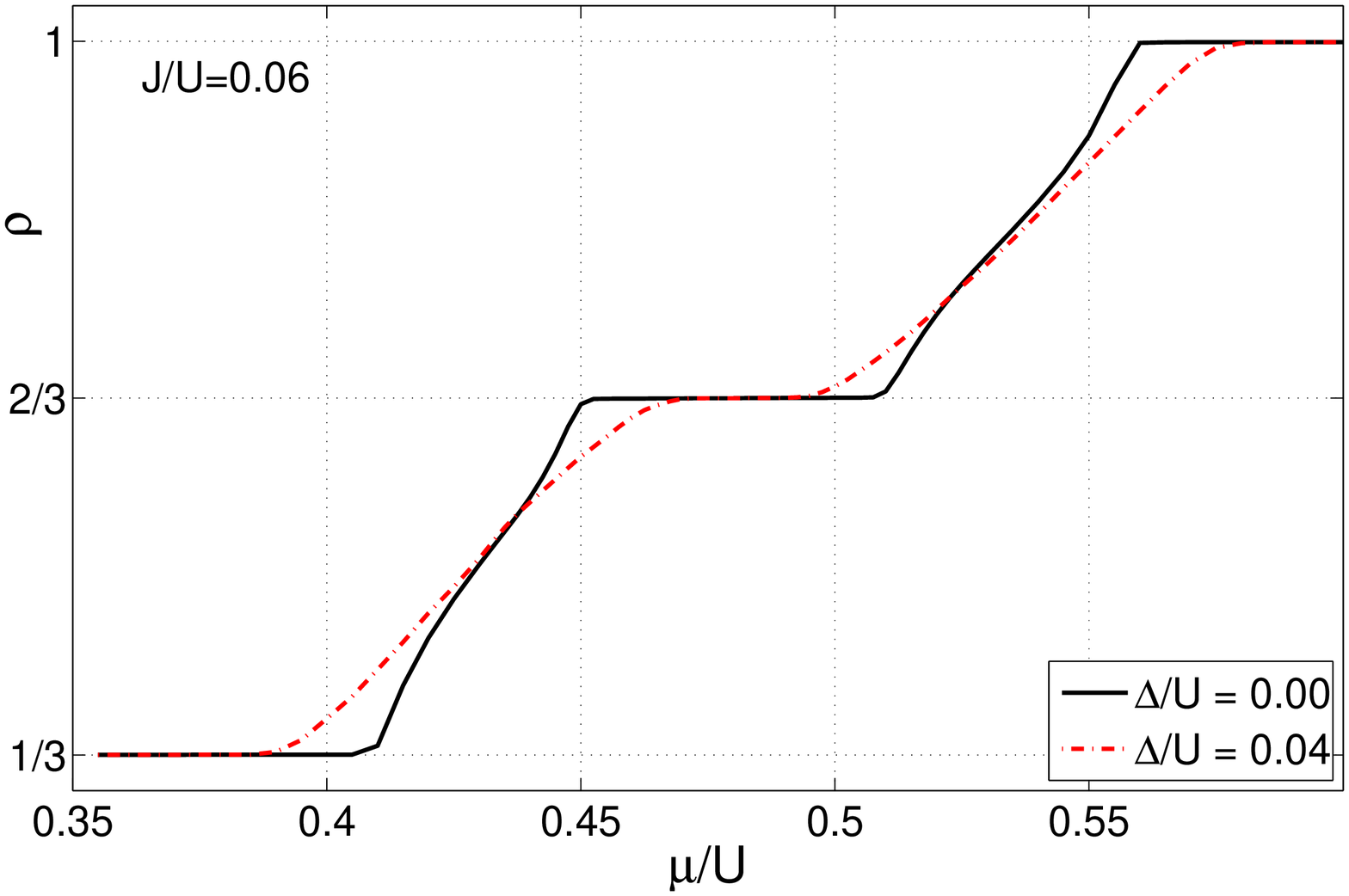,width=\columnwidth}
 \epsfig{file=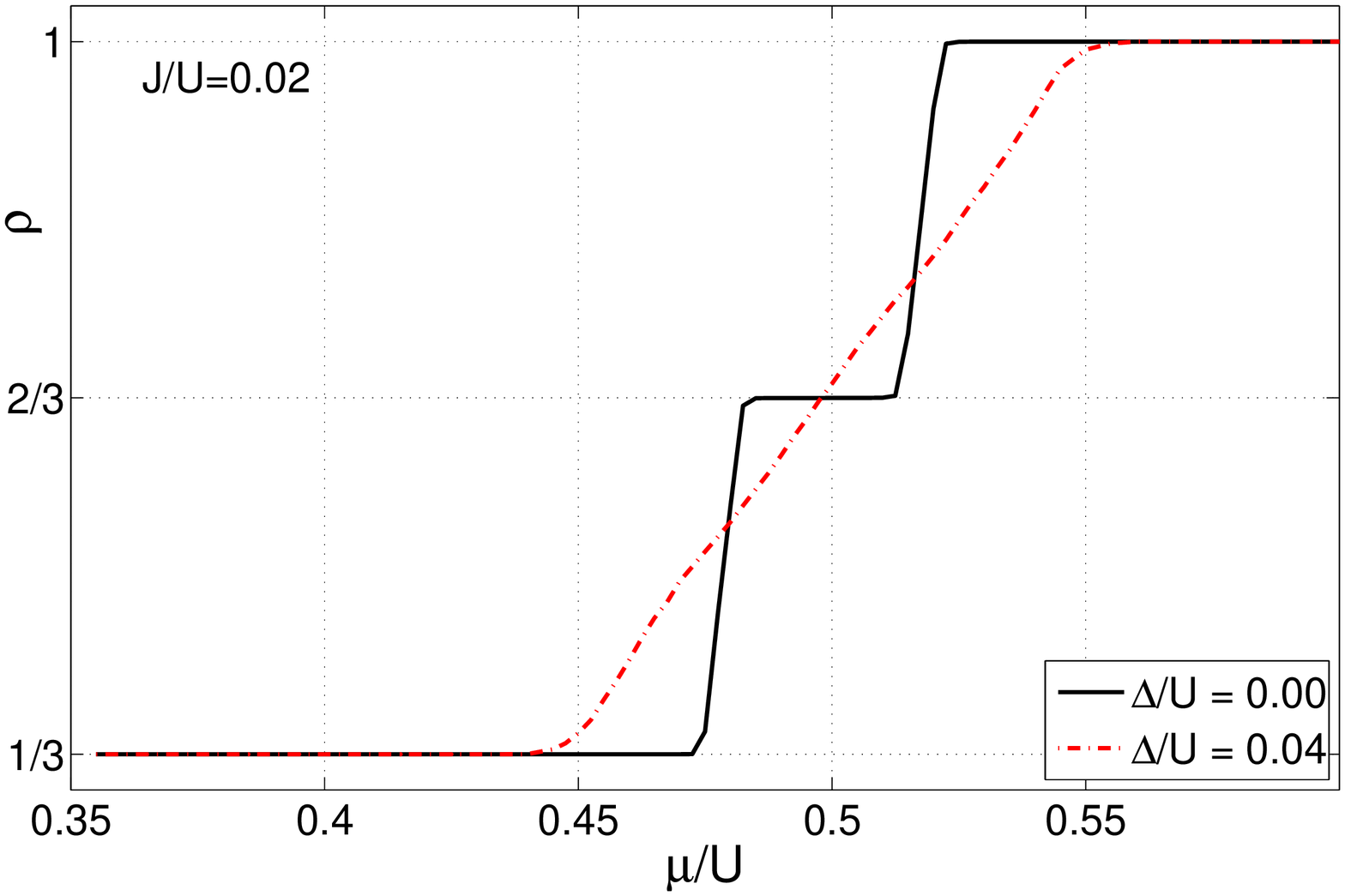,width=\columnwidth}
 \caption{(Color online) Density cut along the horizontal lines in figure \ref{fig:phase_diagram} for the pure (solid line) and the disordered
 ($\Delta/U=0.04$, dashed line) case. Upper plot: cut along the upper line in figure \ref{fig:phase_diagram} at $J/U=0.06$, lower plot: cut 
along the lower line in figure  \ref{fig:phase_diagram} at $J/U=0.02$.  Simulation parameters see figure \ref{fig:phase_diagram_nodisorder}.}
 \label{fig:dichte}
\end{figure}

The decoupling of the loophole insulator from the $J=0$ axis can most easily be seen in a cut parallel to the $\mu$-axis for fixed $J$, showing the average local density as a function of the chemical potential. In the case of small hopping without disorder, this cut shows, beside the expected Mott-lobes at $\rho=\frac13$ and $\rho=1$, an intermediate plateau at filling $\frac23$, corresponding to the loophole phase (Figure \ref{fig:dichte}, lower plot, solid line). For larger hopping (upper plot, solid line), the width of the plateau is slightly increased according to the shape of the loophole in figure \ref{fig:phase_diagram_nodisorder}. In the case of disorder, the plateau for $\rho=\frac23$ vanishes for small hopping (figure \ref{fig:dichte}, lower plot, dashed line). For large hopping (upper plot, dashed line), the incompressible phase survives, however with a largely reduced width compared to the pure case, which shows the decoupling of the loophole from the $J=0$-axis as predicted.\\

\begin{figure}
 \centering
 \epsfig{file=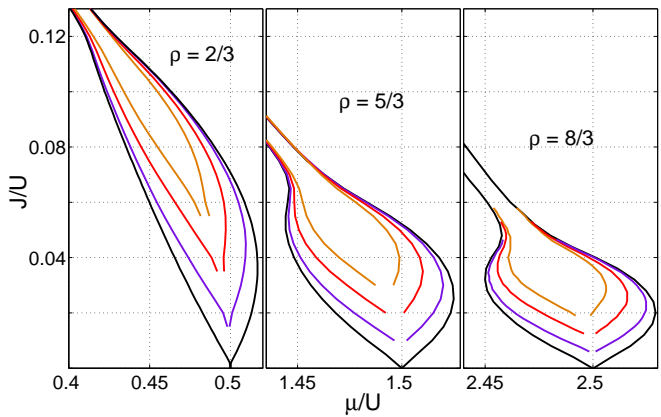,width=\columnwidth}
 \caption{(Color online) Detailed analysis of the first three loophole insulators 
(from left to right: $\rho=2/3$, $\rho=5/3$, $\rho=7/3$) with varying disorder amplitude, \emph{increasing} from the outer to the inner lines 
(black: $\Delta /U=0.00$, magenta: $\Delta /U=0.02$, red: $\Delta /U=0.04$, orange: $\Delta /U=0.06$). For the simulation parameters see figure \ref{fig:phase_diagram_nodisorder} except for $D-1=4$ in the rightmost plot ($\rho=8/3$).}
 \label{fig:loops}
\end{figure}

\begin{figure}
 \centering
  \epsfig{file=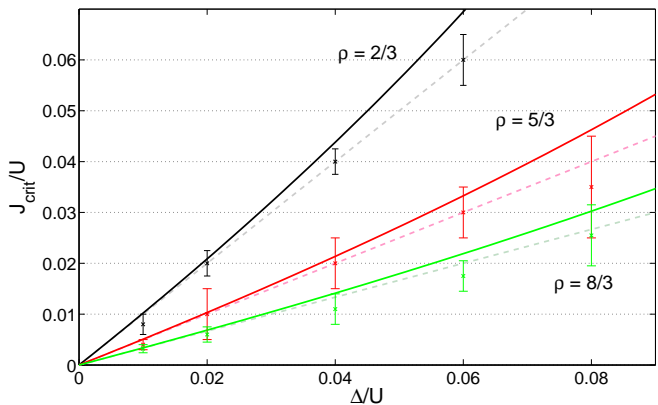,width=\columnwidth}
 \caption{(Color online) Critical point $J_n$ of the onset of the loophole insulating island as a function of the disorder. Solid line: 
analytic prediction from equations (\ref{eq:nis1}) and (\ref{eq:ngt1}), crosses: data with error bars as read from figure \ref{fig:loops} (not all used data is shown in figure \ref{fig:loops}), 
dashed lines: leading order in eq. (\ref{eq:nlin}). From top to bottom: $n=1$, $n=2$, $n=3$.}
 \label{fig:onset}
\end{figure}

In figure \ref{fig:loops} we show the numerical results for the first, second and third loophole for increasing values of the normalised
disorder amplitude. It should be noted that due to the finite number of disorder realisations it is difficult to accurately determine the lower tip of the insulating island in the numerics. In figure \ref{fig:onset} we compare the onset of the insulating loopholes obtained from the analytic two-site model with numerical results. One immediately recognizes two things: First, the onset of the loophole, $J_{\rm crit}$, is a monotonous function of $\Delta$ and second, the higher the filling of the lobe, the earlier the insulating region arises. The numerical value of $J_{\rm crit}$ was obtained by
reading off the values from the numerically determined phase diagram assuming generous error margins \footnote{Calculating the tip of a lobe is in general numerically expensive.}. Taking into account these errors, figure \ref{fig:onset} shows a rather good agreement of the two-site model to the numerics.
However, the analytic prediction tend to be slightly to large compared to the numerics, nevertheless giving the right leading order for small $\Delta$. This is because for larger $\Delta$ the critical hopping gets larger than allowed by the assumption of a decoupled two-site problem. By diagonalising the complete three-site unit cell with periodic boundary-conditions, which gives another but less intuitive approximation, we get a curve for $J_{\rm crit}$ that is below the two-site prediction, but is the same in first order and in better agreement with our numerics.

\section{Summary}
In the present paper we have analyzed the superlattice Bose-Hubbard model with and without disorder. In particular
the case of degenerate potential energies and/or degenerate tunneling rates within the unit cell of the superlattice have
been discussed. Using both, exact numerical methods such as the infinite-size time evolving block decimation (iTEBD) algorithm,
and the density matrix renormalization group (DMRG) we calculated the boundaries of incompressible Mott-insulating
phases. The existence of additional loophole-shaped Mott domains, predicted before, was verified and their numerically
determined phase boundaries compared to other approaches such as the cell strong coupling expansion. A simple effective model was presented that takes the full dynamics within a unit cell into account. The model provides a rather
straight forward explanation for the emergence of loophole Mott domains in the case without disorder. Adding a small
amount of disorder with continuous, bounded distribution lead to a shrinking of the loopholes to Mott insulating islands
with the remarkable feature of a compressible to insulating transition with increasing bosonic hopping. The analytic predictions
for the critical hopping for this transition from the effective model were compared to numerical simulations and found in
very good agreement.

\section*{Acknowledgements}

The authors would like to thank P. Buonsante for providing the CSCPE and QMC data for the
superlattice without disorder. We also thank U. Schollw\"{o}ck for support with the
DMRG calculations. Financial support by the DFG through the SFB-TR 49 and the
GRK 792 is gratefully acknowledged. Also most of the DMRG calculations have been
performed at the John von Neumann-Institut f\"ur Computing, Forschungszentrum J\"ulich.\\

\begin{appendix}\section*{Appendix: The TEBD algorithm and the iTEBD idea}\label{app:tebd}

In the following we give a short summary of the numerical algorithm used in sections \ref{numerics} and \ref{numerics2}. 
The basic idea of the TEBD algorithm emerged from quantum information theory \cite{vidal_part1} and it can be used to simulate one-dimensional quantum computations that involve only a limited amount of entanglement.

Here we want to use it for an imaginary time evolution of the one dimensional Bose Hubbard model. The state of the system can be represented as a matrix product state:
\begin{widetext}\begin{equation}
\vert\Psi\rangle = \sum_{\alpha_1,\alpha_2,\dots\alpha_L = 1}^\chi\sum_{i_1,i_2,\dots i_L = 0}^D
 \Gamma_{\alpha_1}^{[1]i_1}
 \lambda_{\alpha_1}^{[1]}
 \Gamma_{\alpha_1\alpha_2}^{[2]i_2}
 \dots
 \lambda_{\alpha_{L-2}}^{[L-2]}
 \Gamma_{\alpha_{L-2}\alpha_{L-1}}^{[L-1]i_{L-1}}
 \lambda_{\alpha_{L-1}}^{[L-1]}
 \Gamma_{\alpha_{L-1}}^{[L]i_L}
 \vert i_1\dots i_L\rangle. \label{eq:finite_decomposition}
\end{equation}\end{widetext}
Here $D$ is the dimension of the local Hilbert space on a single site and $\chi$ is the number of basis states in the Schmidt decompositions (see below) to be taken into account. $\chi$ is a measure for the maximum entanglement in the system and is assumed not to increase with system size or to increase only very slowly. In our model the number of particles per site is in principle not bounded. But to reduce the numerical effort one can safely set a maximum number of particles $D-1$ allowed per site, since higher occupancies are strongly suppressed due to the on-site interaction. $\vert i_1\dots i_L\rangle$ is the state from the Fock basis, where there are $i_k$ bosons on site $k$.
 
Furthermore we require our matrix product representation to be be in the \emph{canonical} form, i.e. equation (\ref{eq:finite_decomposition}) represents the Schmidt decomposition for any bipartite splitting of the system at the same time. This means for any given $k$, the Schmidt decomposition between sites $k$ and $k+1$ is given by
\begin{equation}
 \vert\Psi\rangle = \sum_{\alpha=1}^\chi \lambda_{\alpha}^{[k]} \vert\Psi_\alpha^{[1\dots k]}\rangle 
\vert\Psi_\alpha^{[k+1\dots L]}\rangle, \label{eq:shortschmidt}
\end{equation}
where the Schmidt coefficients $\lambda^{[k]}_\alpha$ are normalised as
\begin{equation}
 \sum_{\alpha=1}^\chi {\lambda_\alpha^{[k]}}^2 = 1, \label{eq:norm}
\end{equation}
and the $\left\{ \vert\Psi_\alpha^{[1\dots k-1]}\rangle \right\}_\alpha$ $\left(\left\{ \vert\Psi_\alpha^{[k\dots L]}\rangle \right\}_\alpha\right)$ 
form an orthonormal set of states in the subspace of the first $k$ (last $L-k$) sites. By sorting the Schmidt coefficients in an non ascending order for every bond, this makes the representation de facto unique. Explicitly,
\begin{equation}
 \vert\Psi_\alpha^{[1\dots k]}\rangle = \sum_{\substack{\alpha_1,\alpha_2,\dots\alpha_k \\ i_1,i_2,\dots i_k}}
 \Gamma_{\alpha_1}^{[1]i_1}
 \dots
 \lambda_{\alpha_{k-1}}^{[k-1]}
 \Gamma_{\alpha_{k-1} \alpha}^{[k]i_k}
 \vert i_1\dots i_k\rangle,
\end{equation}
where the $\lambda$'s account for all the Schmidt coefficient and the $\Lambda$'s care for the transformation into Fock space at every single site.
Describing an arbitrary state in general requires that the Schmidt number $\chi$ is of the order $D^L$. We will use however a relatively small, constant $\chi$ to avoid exponentially increasing complexity of the numerical problem.
It has been shown in \cite{verstraete}, that this seemingly strong assumption is justified and gives a good approximation for the ground state. 
The latter is related to the fact that the ground state of one-dimensional systems with finite-range interactions has either a constant 
entanglement (for noncritical systems) or the entanglement increases only logarithmically with the size (for critical systems). 
Small values of $\chi$ give usually very good results for local observables, while correlations are only poorly approximated over very large distances.
For the latter the approximation can be improved by choosing a larger $\chi$ proportional to the distance \cite{vidal_itebd}. The amount of coefficients
needed to specify the matrix product state with given fixed $\chi$ is of the order $L\cdot D\cdot\chi^2$ and can be handled numerically in contrast to the $D^L$ coefficients required for representation in the full Fock space.

Expressing the state in a local basis for sites $k$ and $k+1$ only,
\begin{widetext}\begin{equation}
 \vert\Psi\rangle = \sum_{\alpha,\beta,\gamma=1}^\chi
 \sum_{i,j=0}^D
 \lambda_{\alpha}^{[k-1]}
 \Gamma_{\alpha\beta}^{[k]i}
 \lambda_{\beta}^{[k]}
 \Gamma_{\beta\gamma}^{[k+1]j}
 \lambda_{\gamma}^{[k+1]}
 \vert\Psi_\alpha^{[1\dots k-1]}\rangle
 \vert i\rangle \vert j\rangle
 \vert\Psi_\alpha^{[k+2\dots L]}\rangle, \label{eq:longschmidt}
\end{equation}\end{widetext}
we see that applying an operator that involves sites $k$ and $k+1$ only is equivalent to
manipulating the matrices $\Gamma^{[k]}$ and $\Gamma^{[k+1]}$ and the vector $\lambda^{[k]}$ only which is implemented as follows.\\
For reasons of stability we use $A^{[k]i}_{\alpha\beta} := \Gamma_{\alpha\beta}^{[k]i}\lambda_{\beta}^{[k]}$ throughout the algorithm \footnote{The vectors $\lambda^{[k]}$ have all to be kept separately in order to not loose information about the canonical form of the state.}. To shorten the notation we rename $\lambda^{[k-1]}_\alpha\vert\Psi_\alpha^{[1\dots k-1]}\rangle \rightarrow \vert\alpha\rangle$ and $\lambda^{[k+1]}_\gamma\vert\Psi_\gamma^{[k+2\dots L]}\rangle \rightarrow \vert\gamma\rangle$ in (\ref{eq:longschmidt}) giving
\begin{equation}
 \vert\Psi\rangle = \sum_{\substack{\alpha \gamma \\ i j}} 
\sum_{\beta} A^{[k]i}_{\alpha \beta} A^{[k+1]j}_{\beta \gamma}
\frac{1}{\lambda^{[k+1]}_{\gamma}} \vert \alpha i j \gamma \rangle.
\end{equation}
Applying a two site operator $V$ given by the matrix $V^{ij}_{lm}$ then results in
\begin{equation}
 V\vert\Psi\rangle = \vert\widetilde{\Psi}\rangle = \sum_{\substack{\alpha \gamma \\ i j}} 
\underbrace{\sum_{l m \beta} V_{l m}^{i j} A^{[k]l}_{\alpha \beta} A^{[k+1]m}_{\beta \gamma} }
_{T_{\alpha \gamma}^{i j}}
\frac{1}{\lambda^{[k+1]}_{\gamma}} \vert \alpha i j \gamma \rangle.
\end{equation}
 
The objective is now to decompose $T$ into a product of matrices $\widetilde{A}^{[k]i}_{\alpha\beta}\widetilde{A}^{[k+1]j}_{\beta\gamma}$ and to keep the canonical form. The $\left\{\widetilde{A}^{[k+1]j}_{\beta\gamma}\right\}_\beta$ are the eigenvectors of the reduced density matrix
\begin{eqnarray}
\rho^{[k+1\cdots L]} & = & \textrm{Tr}^{[1\cdots k]} \vert\widetilde{\Psi}\rangle \langle\widetilde{\Psi}\vert \\
& = & \sum_{\substack{j_1 j_2 \\ \gamma_1 \gamma_2}} 
\underbrace{ \sum_{i \alpha}{\lambda_{\alpha}^{[k-1]}}^2 T_{\alpha \gamma_1}^{i j_1} \left(T_{\alpha \gamma_2}^{i j_2}\right)^* }_{M_{\gamma_1 \gamma_2}^{j_1 j_2}}
\frac{\vert j_1 \gamma_1 \rangle}{\lambda^{[k+1]}_{\gamma_1}} \frac{\langle j_2 \gamma_2 \vert}{\lambda^{[k+1]}_{\gamma_2}}. \nonumber
\end{eqnarray}
Diagonalising $M$ gives the new $\widetilde{A}^{[k+1]j}_{\beta\gamma}$ as eigenvectors and the new $\left({{\tilde\lambda}^{[k]}_\beta}\right)^2$ as eigenvalues. (Using the $\Gamma$ matrices instead of the $A$ matrices would require a division by $\lambda^{[k+1]}_\gamma$. But $\lambda^{[k+1]}_\gamma$ can be zero if the Schmidt number for this bond is smaller than $\chi$.) In general there are $D\cdot\chi$ nonzero eigenvalues. (This is due to the possible creation of entanglement by $V$.) But we can only keep the $\chi$ biggest of them. Therefor we have to renormalise the new $\lambda^{[k]}$ according to (\ref{eq:norm}). This is necessary anyway if we have a non-unitary $V$ as in the case of an imaginary time evolution. The $\widetilde{A}^{[k]i}_{\alpha\beta}$ are given by $\sum_{j \gamma} \left( \widetilde{A}_{\beta \gamma}^j \right)^* T_{\alpha \gamma}^{ij}$.
 
In order to calculate the ground state of our system (see \cite{vidal_part2}), we divide the Hamiltonian (\ref{eq:BHM}) into two parts 
$\hat{H}_{\rm even}$ and $\hat{H}_{\rm odd}$, where $\hat{H}_{\rm even}$ ($\hat{H}_{\rm odd}$) couples sites $j$ and $j+1$ for even (odd) $j$ only. The local parts of $\hat{H}$ can be distributed between $\hat{H}_{\rm even}$ and $\hat{H}_{\rm odd}$ arbitrarily. The ground state is then given by an imaginary time evolution
\begin{equation}
 \vert\Psi_{\rm ground}\rangle = \lim_{\beta\rightarrow\infty} \frac{e^{-\hat{H}\beta}\vert\Psi_0\rangle}{\parallel e^{-\hat{H}\beta}\vert\Psi_0\rangle\parallel}.
\end{equation}
Here any initial state  $\vert\Psi_0\rangle$ is sufficient, as long as it has a finite overlap with the (yet unknown) ground state. The evolution is implemented by repeatedly applying small time steps $e^{-\hat{H}\varepsilon}$, so called Trotter steps. The norm is conserved in this procedure (see above). So after $T$ steps only the ground state has a reasonable contribution to our state if $\beta = T\cdot\varepsilon$ is much bigger than the inverse of the energy of the first excited state (relative to the ground state energy). In order to write $e^{-\hat{H}\varepsilon}$ as a product of two site operators we use the Suzuki-Trotter decomposition \cite{suzuki}. In first order one can get $e^{-\hat{H}\varepsilon} = e^{-\hat{H}_{\rm even}\varepsilon}e^{-\hat{H}_{\rm odd}\varepsilon} +O(\varepsilon^2)$, in second order $e^{-\hat{H}\varepsilon} = e^{-\frac{\hat{H}_{\rm even}}{2}\varepsilon}e^{-\hat{H}_{\rm odd}\varepsilon}e^{-\frac{\hat{H}_{\rm even}}{2}\varepsilon} +O(\varepsilon^3)$. For higher orders see \cite{suzuki}. Thus we can calculate the ground state by repeatedly applying two-site operators.
 
To calculate expectation values of observables we again take a look at (\ref{eq:longschmidt}). The expectation value of a nearest neighbour observable, say $\langle\Psi\vert\hat{a}^\dagger_k\hat{a}_{k+1}\vert\Psi\rangle$ can be directly calculated because all occurring states are mutually orthogonal and normalised. An $n$th site nearest neighbour observable can be calculated by expressing the state in the local basis for site $k$ to $k+n$ analogous to (\ref{eq:longschmidt}). For non nearest neighbour observables we can use the swap gate to bring the sites of interest together \cite{vidal_part1}.
 
\

\begin{figure}
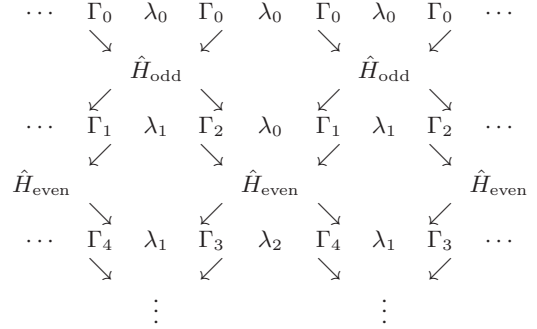

 \centering
 \begin{displaymath}
  \begin{array}{ccccccccc}
   &&&&&&&&\\
   \cdots & \Gamma_0 & \lambda_0 & \Gamma_0 & \lambda_0 & \Gamma_0 & \lambda_0 & \Gamma_0 & \cdots \\
   & \searrow && \swarrow && \searrow && \swarrow &\\
   && \hat{H}_{\rm odd} &&&& \hat{H}_{\rm odd} &&\\
   & \swarrow && \searrow && \swarrow && \searrow &\\
   \cdots & \Gamma_1 & \lambda_1 & \Gamma_2 & \lambda_0 & \Gamma_1 & \lambda_1 & \Gamma_2 & \cdots \\
   & \swarrow && \searrow && \swarrow && \searrow &\\
   \hat{H}_{\rm even} &&&& \hat{H}_{\rm even} &&&& \hat{H}_{\rm even}\\
   & \searrow && \swarrow && \searrow && \swarrow &\\
   \cdots & \Gamma_4 & \lambda_1 & \Gamma_3 & \lambda_2 & \Gamma_4 & \lambda_1 & \Gamma_3 & \cdots \\
   & \searrow && \swarrow && \searrow && \swarrow &\\
   && \vdots &&&& \vdots &&
  \end{array}
 \end{displaymath}
 \caption{(Color online) Symbolic representation of the effect of the TEBD algorithm to a translationally invariant state. The 
uppermost (initial) state is first changed by application of $H_{odd}$, giving new $\Gamma$'s and $\lambda$'s. The second 
step with $H_{even}$ then produces a list of alternating $\Gamma$'s and $\lambda$'s such that only two of them need to be kept in memory. Every further Trotter step preserves this symmetry.}
 \label{fig:itebd}
\end{figure}

A powerful feature of the algorithm is its application to infinite, translationally invariant systems. Suppose a Hamiltonian that has a periodicity of $c$ sites (as (\ref{eq:BHM}) has for a superlattice), restricting to $c=2$ for clarity. The state of an infinite system is a slight modification of (\ref{eq:finite_decomposition}).
\begin{equation}
 \vert\Psi\rangle = \!\!\!\!\!\! \sum_{\substack{\alpha_{-\infty}\dots\alpha_\infty \\ i_{-\infty}\dots i_\infty}} \!\!\!\!\!\!\!\!\!\!
 \dots
 \lambda_{\alpha_{k-1}}^{[k-1]}
 \Gamma_{\alpha_{k-1}\alpha_{k}}^{[k]i_{k}}
 \lambda_{\alpha_{k}}^{[k]}
 \Gamma_{\alpha_k}^{[k+1]i_{k+1}}
 \lambda_{\alpha_{k+1}}^{[k+1]}
 \dots
 \vert i_{-\infty}\dots i_\infty\rangle. \label{eq:infinite_decomposition}
\end{equation}
The imaginary time evolution is started with a translationally invariant state, so all $\Gamma$'s and $\lambda$'s are the same in the beginning. The scheme in figure \ref{fig:itebd} shows, that the $c$-periodicity of the representation is preserved during real or imaginary time evolution. This is because all two-site operations $\hat{H}_{\rm odd}$ and $\hat{H}_{\rm even}$ are the same respectively and are all applied to every other pair of matrices.

So we only have to store two $\Gamma$ matrices and two $\lambda$ vectors. It is even more important that we only have to apply two 
two-site operators per Trotter step.
 
After imaginary time evolution we end up with an $c$-periodic ground state. (That means that expectation values have a periodicity of $c$ sites. Although there can be contributions in (\ref{eq:infinite_decomposition}) from states which have a \emph{nonperiodic} Fock representation. This is a clear distinction from the case of periodic boundary conditions, where not only all expectation values, but also the wave function must be periodic.) This is called the iTEBD algorithm \cite{vidal_itebd}. This means that we can efficiently calculate observables in the thermodynamic limit. If we were using 
DMRG or normal TEBD we would have to simulate large finite systems, which is time consuming, and then extrapolate to $L=\infty$ to get rid of 
finite size effects but introducing additional error.
 
The idea works as well for $c > 2$. If $c$ is odd, we have to choose $2c$ as period, since we need a clear distinction between $\hat{H}_{\rm even}$ and $\hat{H}_{\rm odd}$. In fact we used it in this work for the non periodic Hamiltonian of the disordered superlattice model, thus not saving calculation time (a large value has to be used for $c$ in order to have a sufficiently random disorder) but getting rid of boundary effects.
 
Finally we note that the TEBD algorithm itself is in principle only correct for unitary operations. Non unitary operations were found to destroy the representation in the sense that the Schmidt vectors in (\ref{eq:shortschmidt}) are no longer exactly orthogonal, i.e. the representation is no longer canonical \cite{vidal_itebd2}. Additional steps to conserve orthogonality in the algorithm were proposed in \cite{vidal_tree}. These were not incorporated here, since for small $\varepsilon$ the Trotter steps are quasi orthogonal. Numerical analysis shows, 
that the scalar products of the normalised Schmidt vectors in the resulting ground state are of the order $10^{-3}$.

\end{appendix}


\begin{thebibliography}{99}
 
\bibitem{lit:Jaksch-PRL-1998}
	D.\ Jaksch, C.\ Bruder, J.\ I.\ Cirac, C.\ W.\ Gardiner, and P.\ Zoller, 
	Phys.\ Rev.\ Lett., {\bf 81}, 3108 (1998).
 
\bibitem{lit:Greiner-Nature-2002}
	M.\ Greiner, O.\ Mandel, T.\ Esslinger, T.W.\ H\"{a}nsch, and I.\ Bloch, 
	Nature {\bf 415}, 39 (2002).

\bibitem{lit:santos_prl_2004}
	L. \ Santos, M. \ A. \ Baranov, J. \ I. \ Cirac, H.-U. \ Everts, H. \ Fehrmann and M. \ Lewenstein,
	Phys. Rev. Let.t., {\bf 93}, 030601 (2004)
 
\bibitem{lit:Peil-PRA-2003}
	S. \ Peil, J. \ V. \ Porto, B. \ Laburthe \ Tolra, J. \ M. \ Obrecht, B. \ E. \ King, M. \ Subbotin, S. \ L. \ Rolston, and W. \ D. \ Phillips,
	Phys. Rev. A {\bf 67}, 051603(R) (2003).

\bibitem{lit:Rousseau-PRB-2006}
	V. \ G. \ Rousseau, D. \ P. \ Arovas, M. \ Rigol, F. \ H\'{e}bert, G. \ G. \ Batrouni, and R. \ T. \ Scalettar,
	Phys. Rev. B {\bf 73}, 174516 (2006)

\bibitem{lit:Roth-PRA-2003}
	R.\ Roth, and K.\ Burnett, Phys.\ Rev.\ A, {\bf 68}, 023604 (2003).
 
\bibitem{lit:Buonsante-PRA-2004a}
	P.\ Buonsante and A.\ Vezzani, Phys.\ Rev.\ A {\bf 70}, 033608(R) (2004).

\bibitem{lit:Buonsante-PRA-2004b}
	P.\ Buonsante, V.\ Penna and A.\ Vezzani, Phys.\ Rev.\ A {\bf 70}, 061603(R) (2004).

\bibitem{lit:Buonsante-PRA-2005} 
	P.\ Buonsante and A.\ Vezzani, Phys.\ Rev.\ A {\bf 72}, 013614 (2005).

\bibitem{lit:inguscio_prl_2007}
	L. \ Fallani, J. \ E. \ Lye, V. \ Guarrera, C. \ Fort, and M. \ Inguscio,
	Phys. Rev. Lett. {\bf 98}, 130404 (2007).

\bibitem{vidal_part1}
	G. \ Vidal,
	Phys. Rev. Lett. {\bf 91}, 147902 (2003).

\bibitem{vidal_itebd}
	G. \ Vidal,
	Phys. Rev. Lett. {\bf 98}, 070201 (2007).

\bibitem{lit:Schollwoeck-RMP-2005}
	U.\ Schollw\"ock, Rev.\ Mod.\ Phys.\ {\bf 77}, 000259 (2005).
 
\bibitem{lit:Fisher-PRB-1988}
	M.\ P.\ A.\ Fisher, P.\ B.\ Weichman, G.\ Grinstein, and D. S.\ Fisher, 
	Phys.\ Rev.\ B {\bf 40}, 546 (1989).

\bibitem{lit:Freericks-PRB-1996} 
	J.\ K.\ Freericks and H.\ Monien, 
	Phys.\ Rev.\ B {\bf 53}, 2691 (1996).

\bibitem{lit:Mering-2}
	A.\ Mering and M.\ Fleischhauer,
	Phys. Rev. A, {\bf 77}, 023601 (2008).

\bibitem{verstraete}
	F. \ Verstraete and J. \ I. \ Cirac,
	Phys. Rev. B {\bf 73}, 094423 (2006).
 
\bibitem{suzuki}
	M. \ Suzuki,
	Phys. Lett. A, {\bf 146}, 319 (1990).

\bibitem{vidal_part2}
	G. \ Vidal, 
	Phys. Rev. Lett. {\bf 93}, 040502 (2004).

\bibitem{vidal_itebd2}
	R. \ Orus, G. \ Vidal,
	arXiv:0711.3960 (2007).

\bibitem{vidal_tree}
	Y.-Y. \ Shi, L.-M. \ Duan, G. \ Vidal,
	Phys. Rev. A {\bf 74}, 022320 (2006).

%
%
%
%
%

%
%
%

%

 
%
%
%
%
%
%
%
%
%
%
%
%
%
%
%


\end{thebibliography}
\end{document}